# Physics-informed neural network framework for solving forward and inverse flexoelectric problems

Hyeonbin Moon[1†], Donggeun Park[1†], Jinwook Yeo[1] and Seunghwa Ryu[1*]

[1]Department of Mechanical Engineering, Korea Advanced Institute of Science and Technology (KAIST), 291 Daehak-ro, Yuseong-gu, Daejeon 34141, Republic of Korea

[1†] Hyeonbin Moon and Donggeun Park contributed equally to this work.

[*]Corresponding author e-mail

ryush@kaist.ac.kr (Seunghwa Ryu)

**Abstract**

Flexoelectricity, the coupling between strain gradients and electric polarization, poses significant computational challenges due to its governing fourth-order partial differential equations that require $C^1$-continuous solutions. To address these issues, we propose a physics-informed neural network (PINN) framework grounded in an energy-based formulation that treats both forward and inverse problems within a unified architecture. The forward problem is recast as a saddle-point optimization of the total potential energy, solved via the deep energy method (DEM), which circumvents the direct computation of high-order derivatives. For the inverse problem of identifying unknown flexoelectric coefficients from sparse measurements, we introduce an additional variational loss that enforces stationarity with respect to the electric potential, ensuring robust and stable parameter inference. The framework integrates finite element-based numerical quadrature for stable energy evaluation and employs hard constraints to rigorously enforce boundary conditions. Numerical results for both direct and converse flexoelectric effects show excellent agreement with mixed-FEM solutions, and the inverse model accurately recovers material parameters from limited data. This study establishes a unified, mesh-compatible, and scalable PINN approach for high-order electromechanical problems, offering a promising alternative to traditional simulation techniques.

**Keywords:** Physics-informed neural network; Deep learning; Deep energy method; Flexoelectricity;

## 1. Introduction

Flexoelectricity refers to an electromechanical coupling mechanism in which strain gradients in dielectric materials induce electric polarization, while electric fields, in turn, can produce mechanical strain gradients [1-3]. This effect has received increasing attention in recent years, especially at micro- and nano-scales, where pronounced strain gradients naturally arise due to geometric confinement, material heterogeneity, and interfacial interactions [4-6]. Owing to this size-dependent behavior, flexoelectricity holds great promise for applications in nanoscale sensors, actuators, and energy-harvesting devices [7-10]. The growing interest in these materials has motivated the development of computational models that accurately capture the underlying physics. However, modeling flexoelectric behavior is inherently challenging due to the presence of fourth-order partial differential equations (PDEs) derived from the strain-gradient dependence of the free energy [11-13]. These high-order PDEs require $C^1$- continuous solution fields, which complicates both the variational formulation and the consistent enforcement of boundary conditions. Resolving these difficulties remains a central issue in the computational analysis of flexoelectric phenomena.

To address these challenges, a variety of numerical methods have been developed to accommodate the higher-order continuity requirements intrinsic to strain-gradient theories. Mixed finite element formulations introduce auxiliary variables to reduce the order of the governing equations [14-17], while discontinuous Galerkin methods offer enhanced flexibility in enforcing weak continuity across element boundaries [18-20]. $C^1$ continuous finite elements directly satisfy the required regularity, although their construction and implementation become increasingly complex in nontrivial geometries [21-23]. Isogeometric analysis (IGA), which employs smooth and geometry-conforming basis functions such as B-splines and non-uniform rational B-splines (NURBS), presents a compelling alternative by unifying the geometric and analysis frameworks [24-28]. In addition, meshfree approaches—such as those based on radial basis functions (RBFs)—have been explored for their capacity to satisfy smoothness requirements without reliance on a predefined mesh structure [29, 30]. Furthermore, in cases involving crack tips or material inclusions, analytic solutions have also been developed to complement numerical approaches [31-33].

Recent advances in scientific machine learning have led to the development of physics-informed neural networks (PINN), which offer a new computational paradigm for solving partial differential equations (PDEs) by embedding physical laws directly into the training process of neural networks [34-36]. Leveraging automatic differentiation, PINN incorporate governing equations, boundary conditions, and constitutive relations into a unified loss functional, enabling the approximation of continuous solution fields in a physics-consistent manner. This framework has been successfully applied to various problems in solid mechanics, including linear elasticity, hyperelasticity, elastoplasticity, and polycrystalline plasticity [37-40], as well as to multi-physics systems such as piezoelectricity and electromagnetism [41-43]. However, despite these developments, applications of PINN to flexoelectric problems have received relatively less attention. Although PINN are theoretically capable of handling high-order PDEs, their performance tends to deteriorate with increasing differential order or complexity in boundary conditions [44, 45]. These difficulties are especially pronounced in flexoelectricity, where fourth-order spatial operators and the simultaneous presence of mechanical and electrical boundary conditions pose significant challenges for standard PINN architectures.

In this study, we propose a PINN-based computational framework for modeling flexoelectric materials, capable of addressing the mathematical challenges posed by fourth-order spatial derivatives and complex boundary conditions. Rather than using the strong form of the governing equations, the framework adopts an energy-based formulation that minimizes the total potential energy of the system, commonly referred to as the deep energy method (DEM). This formulation enables systematic handling of higher-order differential operators and facilitates consistent enforcement of boundary conditions. The method is applied to both forward and inverse problems: the forward model predicts displacement and electric potential fields from prescribed boundary conditions (**Fig. 1(a)**), while the inverse model estimates flexoelectric coefficients from limited observation data (**Fig. 1(b)**). Both models are trained using an energy-based optimization scheme, with the forward problem formulated as a saddle-point minimization (**Fig. 1(c)**). To address the limitations of DEM in inverse settings—where pure energy minimization can lead to ill-posed or divergent solutions—the inverse loop incorporates an additional variational loss based on the governing equations (**Fig. 1(d)**). This hybrid strategy enables stable and accurate parameter identification. Numerical experiments confirm that the proposed framework reliably captures the essential features of flexoelectric behavior in both problem types.

The remainder of this paper is structured as follows. **Section 2** presents the governing equations of flexoelectricity. **Section 3** describes the proposed PINN-based methodology. **Section 4** illustrates numerical results for both forward and inverse problems. Finally, **Section 5** concludes the paper.

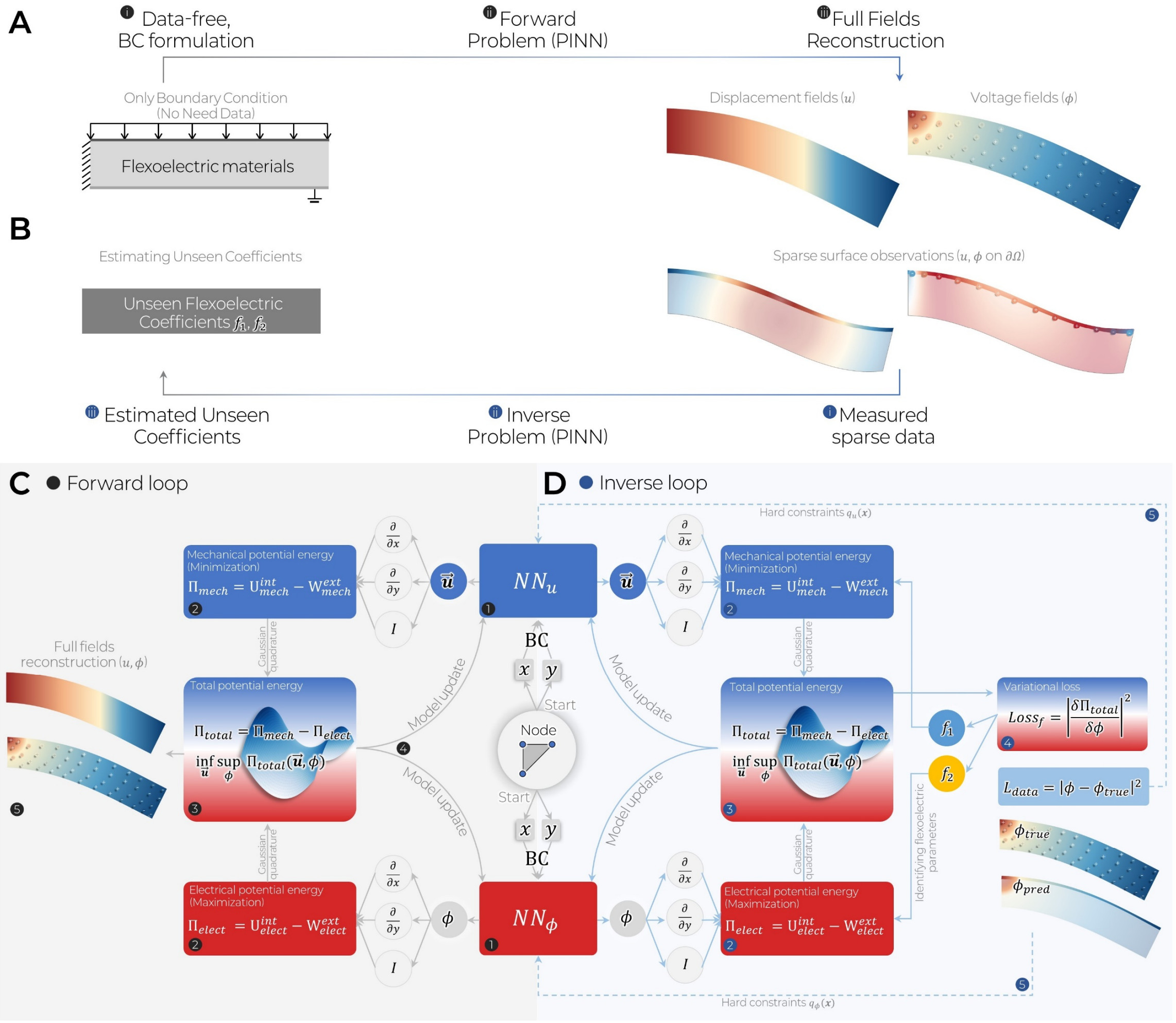

**Fig. 1. Overall schematic of the proposed PINN framework for forward and inverse flexoelectric problems.** (a) Forward: field inference from boundary conditions. (b) Inverse: coefficient identification from measurements. (c) Forward loop: energy-based saddle-point optimization. (d) Inverse loop: coefficient identification with additional variational loss.

## 2. Governing equation

This section introduces the linear flexoelectric theory, originally proposed by Sharma [2], which forms the foundation of the present formulation. The focus is placed on capturing the electromechanical coupling between strain gradients and the electric field. To isolate this interaction, the piezoelectric effect is intentionally excluded from the current framework. Nonetheless, piezoelectric coupling terms—linking strain and electric field—can be readily incorporated into the formulation if needed.

### 2.1) Strong Form

For the linear flexoelectric theory, the governing equations within a bulk domain $\Omega$ subjected to a body force $b_k$ and a free charge density $\rho_0$ are expressed in the strong form as follows:

$$\sigma_{jk,j} - \tau_{ijk,ij} + b_k = 0 \tag{1a}$$

$$D_{i,i} = \rho_0 \tag{1b}$$

Here, $\sigma_{ij}$ , $\tau_{ijk}$ and $D_i$ denote the stress tensor, high-order stress tensor, and electric displacement vector, respectively. Eq. (1a) represents the mechanical equilibrium, while Eq. (1b) corresponds to the Maxwell equation for electrostatics. The constitutive relations for a linear isotropic flexoelectric material are defined as:

$$\sigma_{ij} = \lambda \varepsilon_{kk} \delta_{ij} + 2\mu \varepsilon_{ij} \tag{2a}$$

$$\tau_{ijk} = l^2\left(\lambda \varepsilon_{m\,mi} \delta_{ij} + 2\mu \varepsilon_{jk,i}\right) - f_1 E_i \delta_{jk} - 2 f_2 E_j \delta_k \tag{2b}$$

$$D_i = \kappa E_i + f_1 \varepsilon_{kk,i} + f_2 \varepsilon_{ji,j} \tag{2c}$$

The strain tensor is defined as $\varepsilon_{ij} = \frac{1}{2}(u_{i,j} + u_{j,i})$, where $u_i$ is the displacement field. The electric field is given by $E_i = -\phi_{,i}$, with $\phi$ denoting the electric potential. The material parameters include the Lamé constants $\lambda$ and $\mu$, the internal length scale $l$ associated with strain gradient theory, the dielectric constant $\kappa$, and the flexoelectric coefficients $f_1$ and $f_2$.

The associated boundary conditions are prescribed as follows:

(1) Traction boundary condition

$$\bar{t}_k = \sigma_{jk} n_j - \tau_{ijk\ ,i} n_j - D_j(\tau_{ijk}\ n_i) + (D_l n_l) n_i n_j \tau_{ijk} \quad on\ \partial\Omega_t \tag{3a}$$

(2) High order traction boundary condition

$$\bar{r}_k = \tau_{ijk}\ n_i n_j \quad on\ \partial\Omega_r \tag{3b}$$

(3) Surface charge boundary condition

$$\bar{\omega} = D_i n_i \quad on\ \partial\Omega_\omega \tag{3c}$$

(4) Displacement boundary condition

$$\bar{u}_i = u_i \quad on\ \partial\Omega_u \tag{3d}$$

(5) Normal derivatives boundary condition

$$\bar{v}_i = Du_i = u_{i,j} n_j \quad on\ \partial\Omega_v \tag{3e}$$

(6) Electric potential boundary condition

$$\bar{\phi} = \phi \quad on\ \partial\Omega_\phi \tag{3f}$$

These boundary surfaces satisfy the following conditions:

$$\partial\Omega_t \cup \partial\Omega_u = \partial\Omega, \qquad \partial\Omega_r \cup \partial\Omega_v = \partial\Omega, \qquad \partial\Omega_\omega \cup \partial\Omega_\phi = \partial\Omega$$

$$\partial\Omega_t \cap \partial\Omega_u = \emptyset, \qquad \partial\Omega_r \cap \partial\Omega_v = \emptyset, \qquad \partial\Omega_\omega \cap \partial\Omega_\phi = \emptyset$$

where $\partial\Omega$ denotes the boundary of the domain $\Omega$. Eqs. (3a)–(3c) specify the Neumann boundary conditions, while Eqs. (3d)–(3f) define Dirichlet boundary conditions.

## 2.2) Variational form and energy-based formulation

The governing equations in Eqs. (1a) and (1b), together with the boundary conditions in Eqs. (3a)–(3f), can be equivalently reformulated in variational form based on the principle of virtual work:

$$\int_{\Omega} \sigma_{ij}\, \delta\varepsilon_{ij} + \tau_{ijk}\, \delta\varepsilon_{jk,i} dV - \int_{\Omega} b_i \delta u_i \,\mathrm{dV} - \int_{\Omega_t} \bar{t}_i \delta u_i dS - \int_{\Omega_r} \bar{r}_i \delta v_i dS = 0 \tag{4a}$$

$$\int_{\Omega} D_i \delta E_i \, dV - \int_{\Omega} \rho_0 \delta\phi \,\mathrm{dV} - \int_{\Omega_\omega} \bar{\omega} \delta\phi dS = 0 \tag{4b}$$

Here, $\delta u_i$ and $\delta\phi$ are admissible test functions that satisfy the corresponding Dirichlet boundary conditions. This variational form reduces the differential order of the governing equations and provides a foundation for energy-based formulations. Based on this variational formulation, the problem can alternatively be expressed as a saddle-point problem by seeking a stationary point of the total potential energy $\Pi_{total}$, from which the displacement and electric potential fields are obtained as follows:

$$\inf_{\boldsymbol{u}} \sup_{\phi} \Pi_{total} \tag{5a}$$

$$\Pi_{total} = \Pi_{mech} - \Pi_{elect} \tag{5b}$$

The mechanical and electrical potential energies are defined as:

$$\Pi_{mech} = \int_{\Omega} \frac{1}{2} \sigma_{ij}\, \varepsilon_{ij}\, dV + \frac{1}{2} \tau_{ijk}\, \varepsilon_{jk,i} dV - \int_{\Omega} b_i u_i \, dV - \int_{\Omega_t} \bar{t}_i u_i dS - \int_{\Omega_r} \bar{r}_i v_i dS \tag{5c}$$

$$\Pi_{elect} = \int_{\Omega} \frac{1}{2} D_i E_i \, dV - \int_{\Omega} \rho_0 \delta\phi \, dV - \int_{\Omega_\omega} \bar{\omega} \delta\phi dS \tag{5d}$$

In the context of PINN, the governing equations can be incorporated using one of three

formulation strategies:

(i) strong form, using Eqs. (1a)–(1b), referred to as naïve PINN [34-36]

(ii) variational form, using Eqs. (4a)–(4b), referred to as variational PINN [46, 47]

(iii) potential energy–based formulation, using Eqs. (5a)–(5d), referred to as the deep energy method [40, 48]

The strong form requires fourth-order derivatives in the loss function, which can lead to instability and poor convergence during training. The variational form alleviates this issue by lowering the differentiation order, but introduces complications in evaluating variations such as $\delta \mathrm{u}$, particularly for strain gradient terms. To address these limitations, we adopt the potential energy–based formulation, referred to as the deep energy method. This approach reduces the overall differentiation order by one compared to the strong form and facilitates the imposition of the boundary conditions defined in Eqs. (3a)–(3f) within a variational framework. The implementation details are presented in **Section 3**, where we utilize a finite element–based integration scheme to numerically evaluate the total energy functional in Eq. (5), enabling stable and efficient training of the PINN model.

## 3. Methodology

This section presents a PINN framework based on energy principles for solving both forward and inverse problems in flexoelectric materials. The forward problem is defined as predicting the displacement and electric potential fields by solving the governing partial differential equations (PDEs) under prescribed boundary conditions. The inverse problem, in contrast, aims to identify unknown flexoelectric coefficients from partial observations of the solution fields—specifically, from electric potential measurements in the present formulation.

### 3.1) Neural Networks

The proposed PINN framework employs two separate feedforward deep neural networks: one for predicting the displacement field ($NN_u$), and the other for the electric potential field ($NN_\phi$). Each network takes the spatial coordinates $\boldsymbol{x} = \{x, y\}$ as input. The displacement network outputs $\boldsymbol{u}^{NN} = \{u_x^{NN}, u_y^{NN}\}$, while the electric potential network outputs $\phi^{NN}$. The structure of each network is defined as follows:

$$\boldsymbol{z_0} = \{x, y\} \tag{6a}$$

$$\boldsymbol{z_i} = \sigma(\boldsymbol{W_i}\boldsymbol{z_{i-1}} + \boldsymbol{b_i}) \quad wt\,h \quad i \in \{1, \dots, N_{layer}\} \tag{6b}$$

$$NN(\boldsymbol{x}) = \boldsymbol{W_{N+1}}\boldsymbol{z_N} + \boldsymbol{b_{N+1}} \tag{6c}$$

Here, $\boldsymbol{z_i}$ denotes the output of the $i$-th hidden layer, and $\boldsymbol{W_i}$, $\boldsymbol{b_i}$ are the corresponding weight matrix and bias vector, respectively. represent the corresponding weight matrix and bias vector, respectively. These parameters are optimized during training. The nonlinear activation function is denoted by $\sigma$.

To enforce Dirichlet boundary conditions, hard constraints are incorporated through masking and auxiliary functions:

$$\boldsymbol{u}^{NN} = p_u(\boldsymbol{x})NN_u(\boldsymbol{x}) + q_u(\boldsymbol{x}) \tag{7a}$$

$$\phi^{NN} = p_\phi(\boldsymbol{x})NN_\phi(\boldsymbol{x}) + q_\phi(\boldsymbol{x}) \tag{7b}$$

In these expressions, $p_u(\boldsymbol{x})$ and $p_\phi(\boldsymbol{x})$ are masking functions that vanish on the Dirichlet boundaries, ensuring that the neural network output does not influence the solution at those locations. The functions $q_u(\boldsymbol{x})$ and $q_\phi(\boldsymbol{x})$ are auxiliary functions specifically constructed to satisfy the given Dirichlet boundary conditions. These functions are problem-dependent and must be defined accordingly for each application.

### 3.2) Deep energy method for forward problem

The deep energy method (DEM) formulates the training of neural networks by extremizing the total potential energy functional defined in Eqs. (5a)–(5d). To evaluate this energy, it is necessary to compute not only the displacement and electric potential fields but also the derived quantities—namely, strain, strain gradient, and electric field—followed by numerical integration over the domain. The computational domain is discretized using a triangular mesh, and linear shape functions $\boldsymbol{N}$, as typically employed in finite element methods, are used to facilitate numerical differentiation and integration via Gaussian quadrature. The neural networks $NN_u$ and $NN_\phi$ infer the displacement $\boldsymbol{u}^{NN}$ and electric potential $\phi^{NN}$ at mesh nodes. The strain $\boldsymbol{\varepsilon}^{NN}$ is then computed via automatic differentiation.

At each integration point, strain gradients and electric fields are reconstructed using shape function derivatives—represented through the $\boldsymbol{B}$ matrices—as illustrated in **Fig. 2**. The corresponding constitutive variables, including the Cauchy stress $\boldsymbol{\sigma}$, high-order stress $\boldsymbol{\tau}$, and electric displacement **D**, are then evaluated via the relations in Eqs. (2a)–(2c). The mechanical and electrical potential energies, $\Pi_{mech}$ and $\Pi_{elect}$, are integrated using Gaussian quadrature to obtain the total potential energy $\Pi_{total}$. To solve the saddle-point problem in Eq. (5a), the displacement network $NN_u$ is trained to minimize $\Pi_{total}$, while the electric potential network $NN_\phi$ is trained to maximize it.

For enhanced numerical stability during training, we employ the Adam optimizer in combination with an exponential moving average (EMA) update scheme [49, 50], which is known to improve convergence in min–max optimization settings. In this scheme, the model parameters are iteratively updated by applying EMA to the output of the Adam optimizer. The update at iteration $t$ t is given by:

$$\theta_{EMA}^{(t)} = \beta \cdot \theta_{EMA}^{(t-1)} + (1-\beta) \cdot Adam\left(\theta_{EMA}^{(t-1)}\right) \tag{8}$$

where $\theta$ denotes the neural network parameters to be updated, and $\beta \in [0,1)$ is the decay rate that weights the contribution of previous parameters.

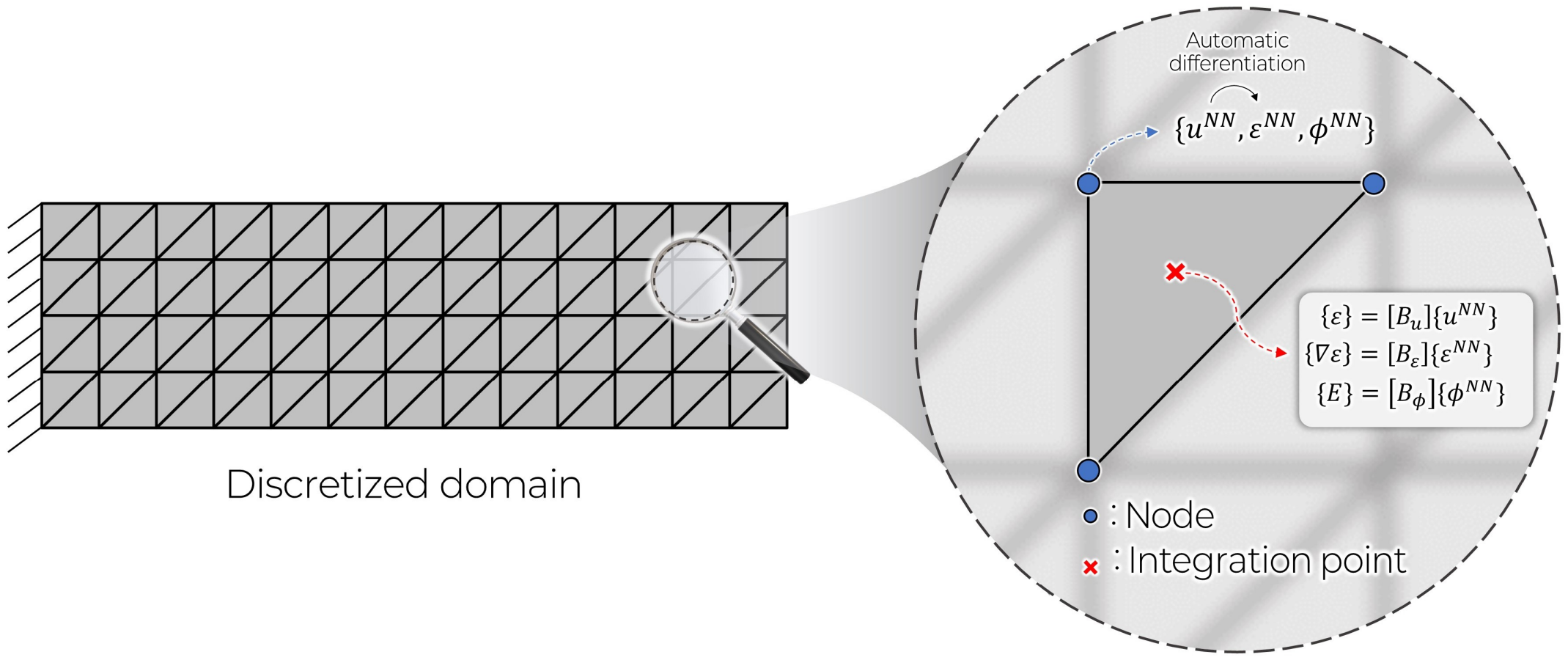

**Fig. 2. Finite element–based differentiation and integration of PINN outputs over a discretized domain**

### 3.3) Variational loss for inverse problem

While the deep energy method has been shown to be effective for solving forward problems with known material parameters, it is not directly applicable to inverse problems where the material properties—such as the flexoelectric coefficients $f_1$ and $f_2$—are treated as unknowns. This limitation stems from the fact that the loss function in the deep energy framework is based on extremizing the total potential energy, as formulated in Eq. (5a). This condition does not enforce strict equality, and when material parameters are included as trainable variables, naive optimization often results in divergence of both the parameters and the predicted electric potential.

To overcome this issue, we introduce an additional loss term based on the variation of the total potential energy with respect to the electric potential. Specifically, we enforce the first-order stationarity condition by minimizing the residual of the following expression:

$$Loss_f = \left|\frac{\delta\Pi_{total}}{\delta\phi}\right|^2 = \left|\int_\Omega \boldsymbol{B}_\phi \boldsymbol{D}\, dV - \int_\Omega \boldsymbol{N}\rho_0\, \mathrm{dV} - \int_{\Omega_\omega} \boldsymbol{N}\bar{\omega} dS\right|^2 \tag{9}$$

This equality-based loss stabilizes the optimization process and enables reliable identification of the flexoelectric coefficients. Notably, while the displacement and electric potential networks ($NN_u$ and $NN_\phi$) are trained using the same energy-based formulation as in the forward problem, the flexoelectric parameters $f_1$ and $f_2$ are exclusively optimized through this variational loss. This decoupled approach prevents divergence of the energy functional and promotes stable convergence. Notably, we do not employ the variation with respect to displacement, as it involves second-order derivatives (i.e., strain gradients), whose differential relationship with strain is not straightforward to preserve under standard finite element.

In this study, we assume that electric potential measurements are available at selected spatial locations. These values are treated as Dirichlet-type constraints and enforced via hard constraints using the masking and auxiliary functions defined in Eqs. (7a)–(7b). This approach ensures exact satisfaction of the observed data without introducing a separate data loss term. Additionally, the use of hard constraints eliminates the need for manual weighting between physics and data losses, thereby simplifying the optimization and enhancing robustness in data-limited, parameter-sensitive inverse problems.

### 3.4) Non-dimensionalization

To enhance numerical stability and enable scale-independent analysis, all governing equations and material parameters are expressed in nondimensional form using appropriate reference quantities. Specifically, a reference length $L_{ref}$ , stiffness $C_{ref}$ , and dielectric constant $\kappa_{ref}$ are introduced to normalize geometric, mechanical, and electrical quantities. The nondimensional quantities are defined as follows:

$$\boldsymbol{x}^* = \frac{\boldsymbol{x}}{L_{ref}}, \qquad \boldsymbol{u}^* = \frac{\boldsymbol{u}}{L_{ref}}, \qquad l^* = \frac{l}{L_{ref}},$$

$$\boldsymbol{\sigma}^* = \frac{\boldsymbol{\sigma}}{C_{ref}}, \qquad \boldsymbol{\tau}^* = \frac{\boldsymbol{\tau}}{C_{ref}\, L_{ref}}, \qquad \lambda^* = \frac{\lambda}{C_{ref}}, \qquad \mu^* = \frac{\mu}{C_{ref}},$$

$$\kappa^* = \frac{\kappa}{\kappa_{ref}}, \qquad \phi^* = \frac{\phi}{L_{ref}}\sqrt{\frac{\kappa_{ref}}{C_{ref}}}, \qquad \rho_0^* = \frac{\rho_0}{\sqrt{\kappa_{ref}\, C_{ref}}\,/L_{ref}^2}$$

$$f_1^* = \frac{f_1}{L_{ref}\sqrt{\kappa_{ref}\, C_{ref}}}, \qquad f_2^* = \frac{f_2}{L_{ref}\sqrt{\kappa_{ref}\, C_{ref}}}$$

These nondimensional variables are consistently used throughout the formulation, including in the governing equations, boundary conditions, and loss function definitions. Unless explicitly stated otherwise, all quantities presented in the remainder of the paper are nondimensional.

## 4. Numerical experiments

Numerical experiments were performed for both forward and inverse problems using a two-dimensional flexoelectric model. The PINN architecture consisted of fully connected feedforward neural networks with six hidden layers and 64 neurons per layer, utilizing the hyperbolic tangent (tanh) activation function. Model training was conducted using the Adam optimizer in combination with an EMA scheme, with a learning rate of 1e-3 and a decay rate of $\beta = 0.95$. All implementations were carried out in the PyTorch framework.

For the forward problem, the computational domain was discretized using 3,276 nodes, while 6,552 nodes were used for the inverse problem to accommodate additional spatial resolution requirements. The predictions obtained from the PINN framework were validated against reference solutions computed using the mixed finite element method (mixed FEM) [14-17]. The material properties and reference values adopted for nondimensionalization are summarized in **Table 1**. The explicit forms of the masking and auxiliary functions employed to enforce Dirichlet boundary conditions via hard constraints are provided in **Supplementary Note 1**.

**Table 1.** List of material properties

| Parameter | Notation | Unit | Value |
|---|---|---|---|
| Lamé constant | $\lambda$ | $N/m^2$ | 8.0192e+10 |
| - | $\mu$ | $N/m^2$ | 5.3462e+10 |
| Strain gradient length scale | $l$ | m | 4e-7 |
| Dielectric constant | $\kappa$ | $C/V \cdot m$ | 1e-9 |
| Flexoelectric coefficient | $f_1$ | $C/m$ | 1e-6 |
| - | $f_2$ | $C/m$ | 1e-6 |
| Reference length | $L_{ref}$ | m | 1e-6 |
| Reference stiffness | $C_{ref}$ | $N/m^2$ | 139e+9 |
| Reference dielectric constant | $\kappa_{ref}$ | $C/V \cdot m$ | 1e-9 |

### 4.1) Forward problem

To assess the effectiveness of the proposed deep energy method in solving forward flexoelectric problems, two benchmark configurations were investigated, as illustrated in **Fig. 3**. The first configuration, shown in **Fig. 3(a)**, corresponds to the direct flexoelectric effect, where mechanical loading generates a strain gradient that, through electromechanical coupling, induces an electric potential. The second configuration, shown in **Fig. 3(b)**, represents the converse flexoelectric effect, in which an externally applied electric field induces mechanical deformation in the absence of mechanical loading.

In both cases, the structure is modeled as a cantilevered beam, with the left edge fully clamped and the bottom edge electrically grounded. In the direct case, a uniform pressure is applied along the top edge, leading to bending and a resulting strain gradient that induce an electric potential. In the converse case, an electric potential is applied to the top edge, and the resulting electric field produces a strain gradient that induces mechanical deformation, even in the absence of external mechanical loading

**Fig. 4** presents the results of the forward problem for the direct flexoelectric effect. A uniform mechanical load is applied to the top edge of the beam, generating strain gradients that induce an electric potential which characterizes the direct flexoelectric response. As shown in **Fig. 4(a)**, the mechanical and electrical components of the total potential energy converge stably after approximately 10,000 epochs. **Fig. 4(b)** shows that the PINN accurately predicts displacement, electric potential, and strain fields in close agreement with FEM results. Notably, the induced electric potential distribution is well reproduced, despite being generated purely from mechanical loading. In particular, the predicted electric field—computed as the gradient of the potential—closely matches the FEM solution even near the left corner, where strain gradients are most pronounced. This indicates that the PINN effectively captures the direct electromechanical coupling, especially in regions of strong spatial variation. The absolute error in the electric potential field remains below the order of $10^{-4}$, and the strain field shows similar accuracy levels across most of the domain, except near the left corner where localized discrepancies appear. This region corresponds to a theoretical singularity, where stress, strain, and their gradients diverge, leading to sharp electric field localization.

Similarly, **Fig. 5** presents the results of the forward problem for the converse flexoelectric effect. In this case, an electric potential difference is applied across the top and bottom edges, generating an electric field that induces mechanical deformation and strain, even without external mechanical loading. **Fig. 5(a)** shows that the mechanical and electrical components of the total potential energy converge smoothly during training, indicating stable learning behavior. As shown in **Fig. 5(b)**, the PINN accurately captures the displacement and strain fields, with results closely matching the FEM reference. Notably, the model successfully reproduces the strain response driven purely by the electric field, demonstrating its ability to resolve converse electromechanical coupling. The predicted strain field shows good agreement across most of the domain, with absolute errors generally within the order of $10^{-4}$. As in the direct flexoelectric case, slightly larger discrepancies are observed near the left corner, where localized field concentration occurs due to geometric and boundary effects.

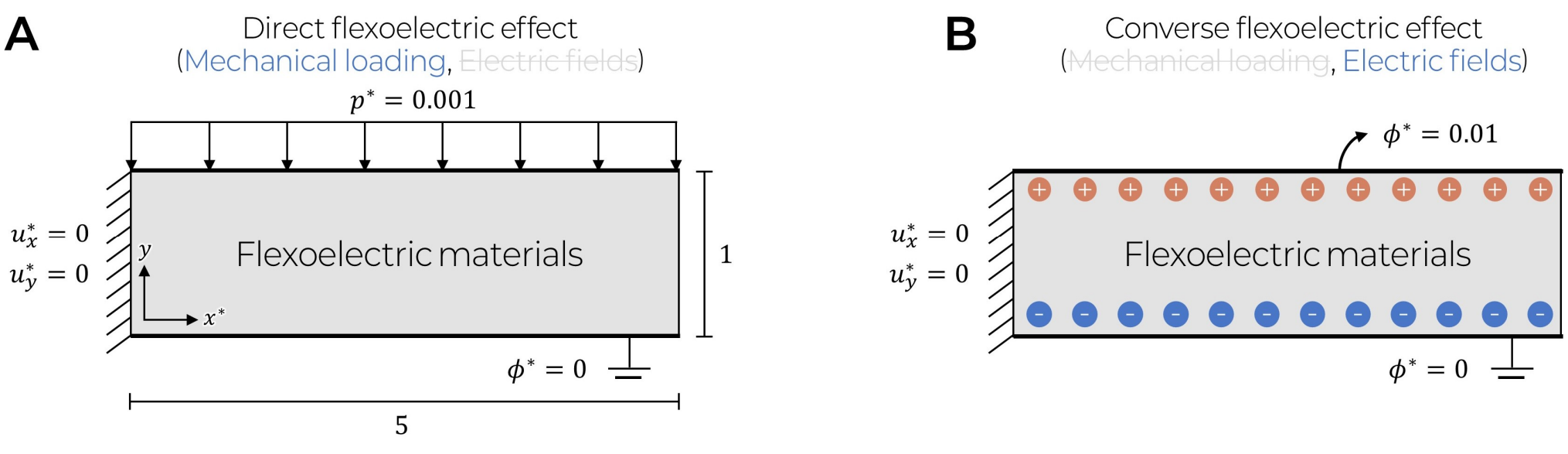

**Fig. 3 Boundary conditions for forward flexoelectric problems.** (a) Direct flexoelectric effect: mechanical loading induces polarization via strain gradients. (b) Converse flexoelectric effect: applied electric field generates mechanical deformation.

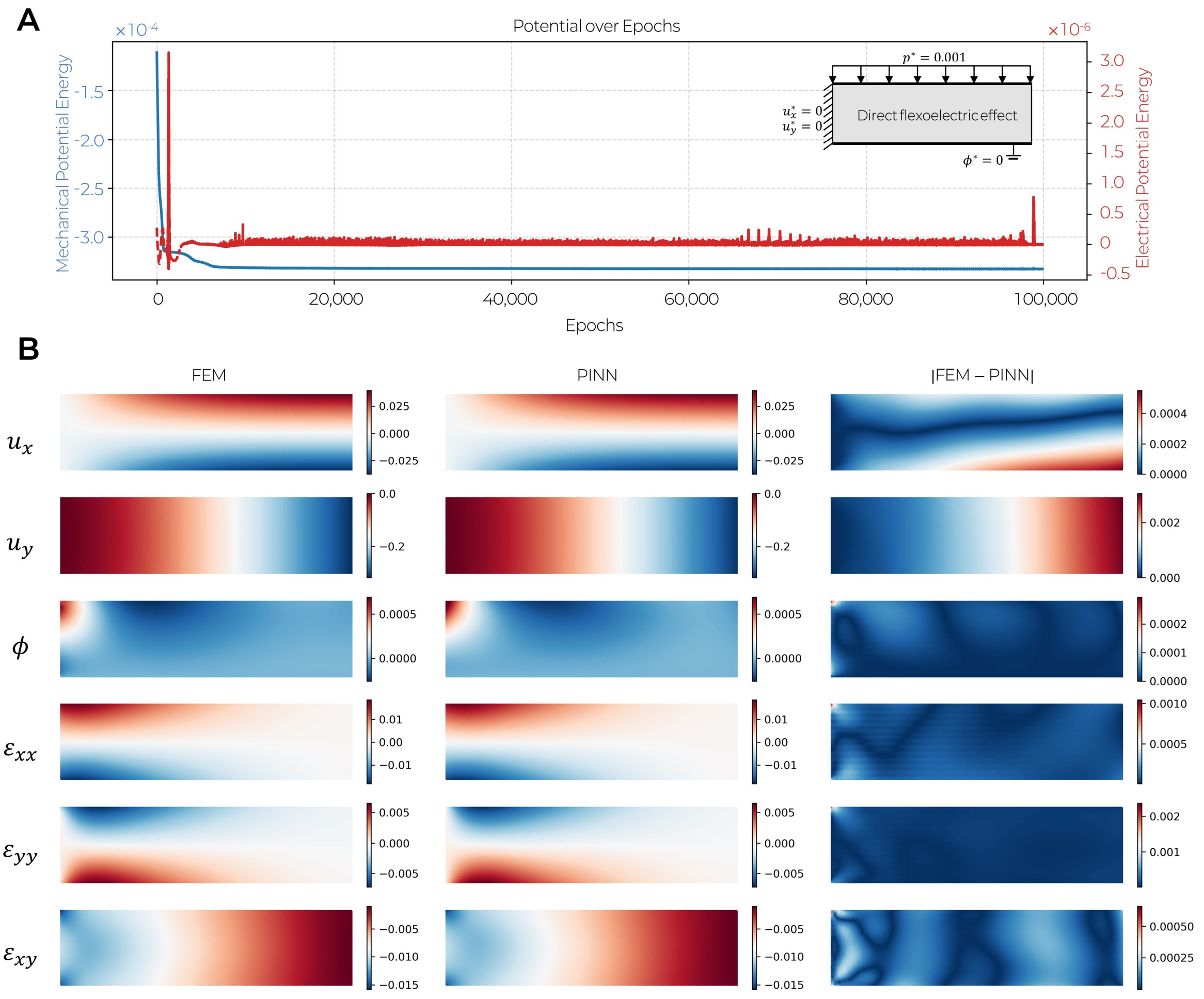

**Fig. 4. Results of the forward problem for the direct flexoelectric effect.** (a) Evolution of mechanical and electrical potential energy during PINN training. (b) Comparison of predicted displacement, electric potential, and strain fields with FEM reference.

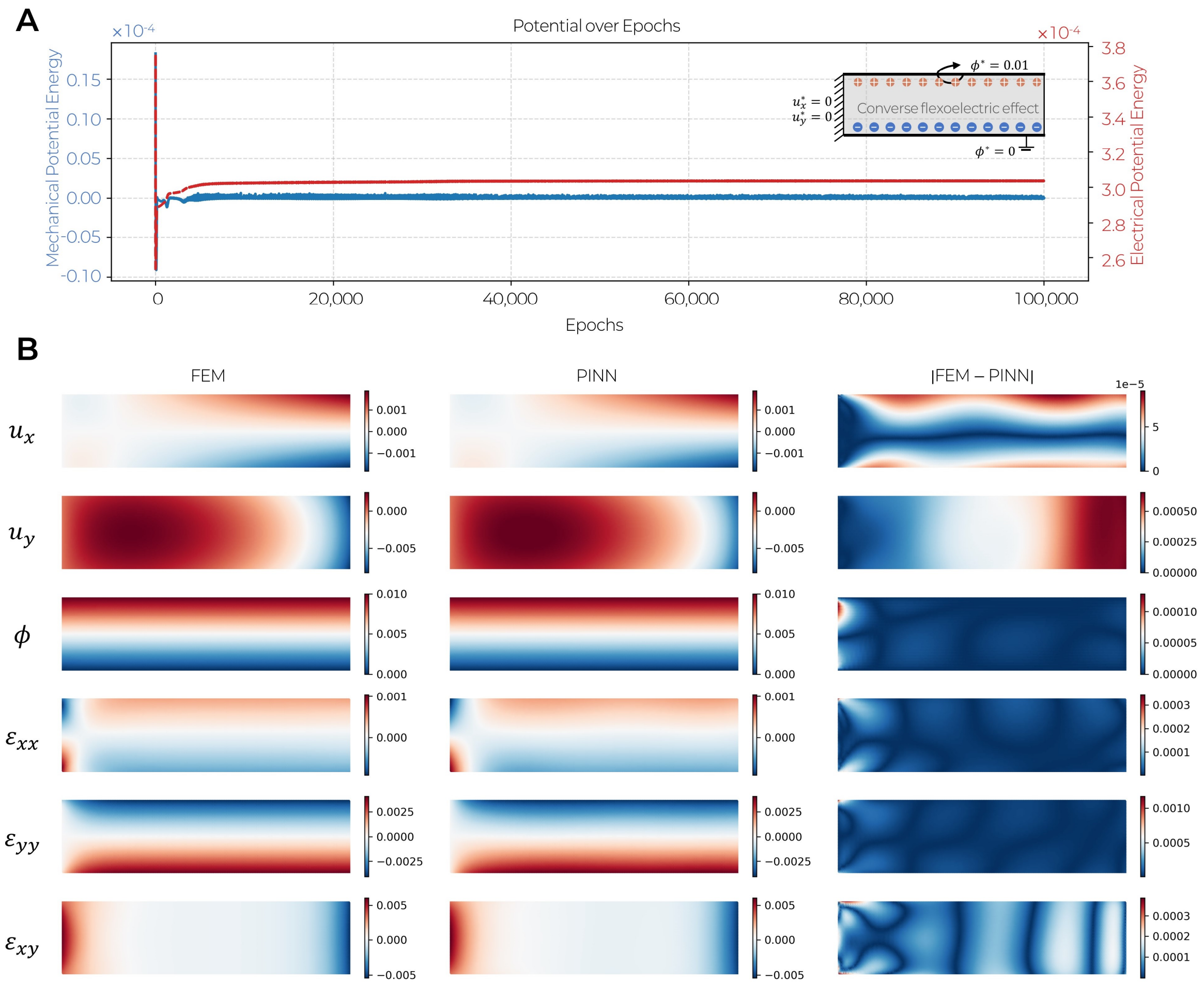

**Fig. 5. Results of the forward problem for the converse flexoelectric effect.** (a) Evolution of mechanical and electrical potential energy during PINN training. (b) Comparison of predicted displacement, electric potential, and strain fields with FEM reference.

### 4.2) Inverse problem

To demonstrate the proposed framework in an inverse setting, we consider a problem in which the flexoelectric coefficients are estimated concurrently with the reconstruction of displacement and electric potential fields. This inverse formulation builds upon the deep energy method augmented with the variational loss introduced in **Section 3.3**, where stationarity with respect to the electric potential is enforced to stabilize training and avoid parameter divergence. As illustrated in **Fig. 6(a)**, the problem setup involves a beam clamped at both lateral edges and subjected to a uniform pressure along the bottom edge. The electric potential distribution along the top edge (**Fig. 6(b)**) is used as the measurement input for the inverse process. While FEM-generated data are used in this study for validation purposes, such measurements could also be obtained experimentally using techniques such as atomic force microscopy (AFM). Following the strategy outlined in **Section 3.3**, the measurement data are imposed as Dirichlet boundary conditions through hard constraints implemented via masking and auxiliary functions (see **Supplementary Note 1**). This approach removes the need for an explicit data loss term, simplifying the overall loss formulation and ensuring exact satisfaction of the observed values.

The current configuration is designed to ensure the mathematical well-posedness of the inverse problem. In practical scenarios, while the boundary conditions, geometry, and measurement locations may vary, the proposed framework remains applicable as long as the inverse problem is well-posed.

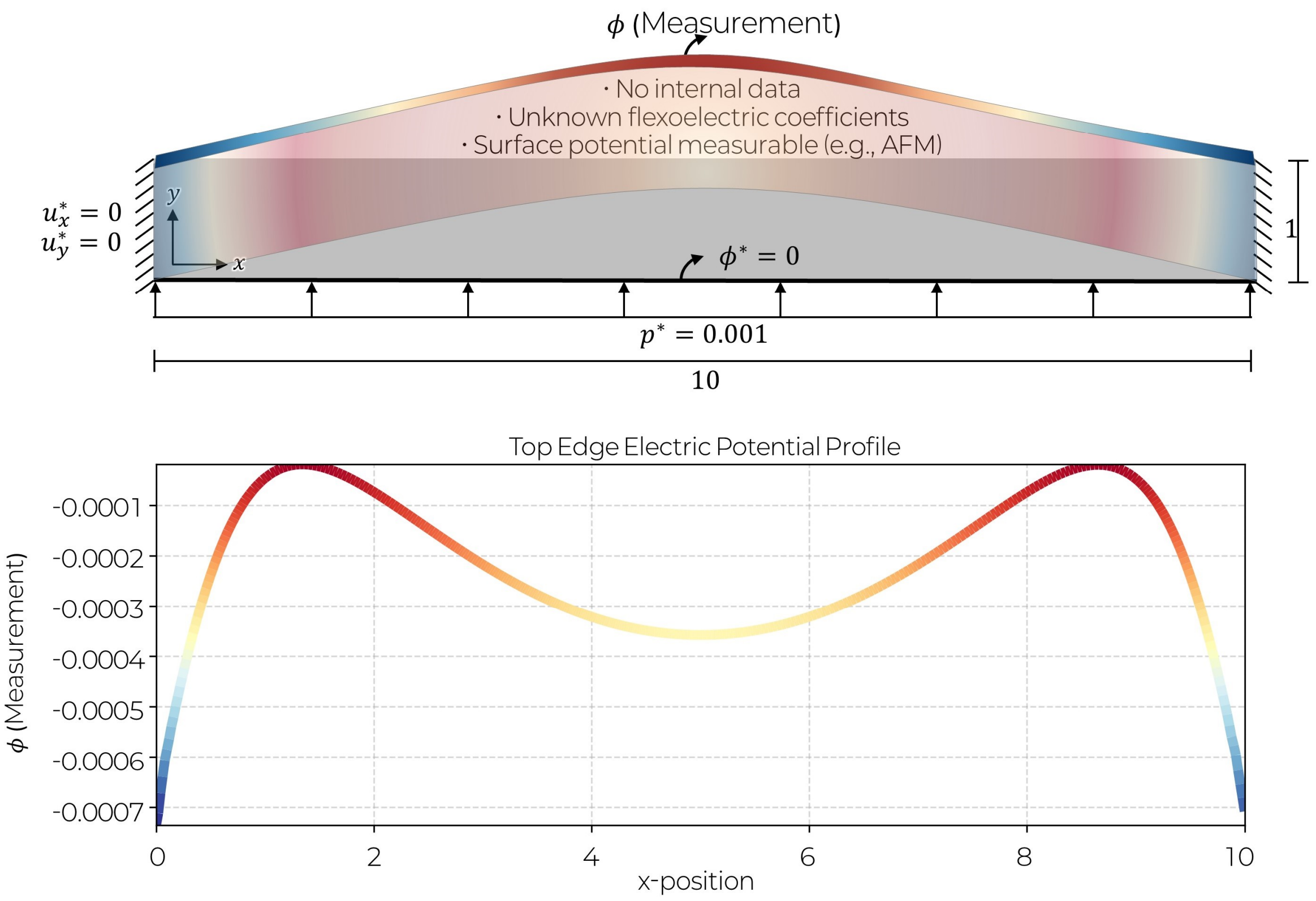

**Fig. 6. Inverse problem setup for flexoelectric parameter identification.** (a) Boundary conditions with electric potential measured along the top surface. (b) Measured electric potential profile prescribed as a Dirichlet condition.

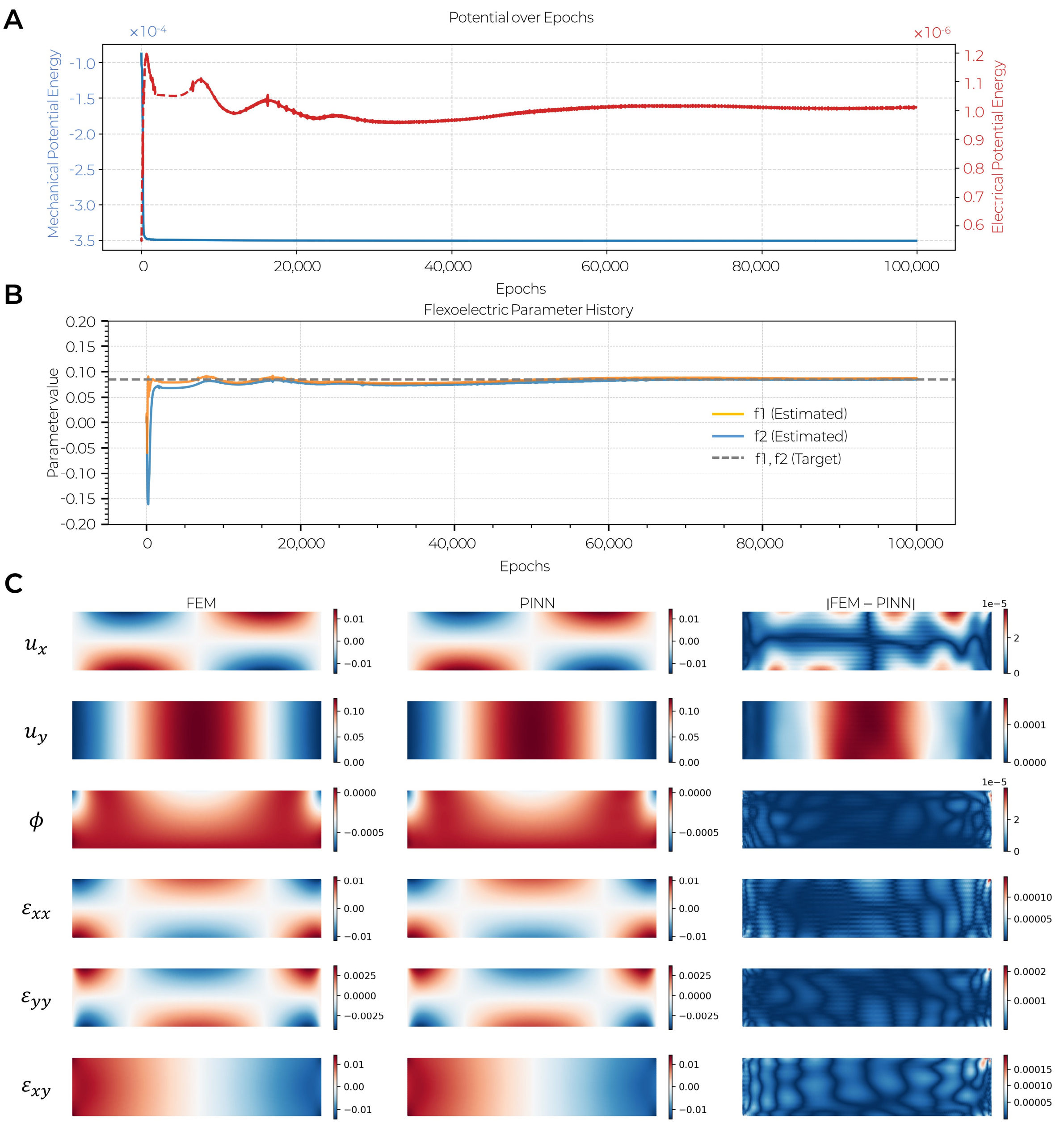

**Fig. 7 Results of the inverse problem for flexoelectric coefficient identification.** (a) Evolution of mechanical and electrical potential energy during training. (c) Training history of the estimated flexoelectric parameters. (c) Comparison of predicted displacement, electric potential, and strain fields with FEM reference

**Fig. 7** validate the proposed PINN framework for solving the inverse flexoelectric problem. In **Fig. 7(a)**, both the mechanical and electrical potential energies converge smoothly to their respective extrema, reflecting stable and consistent optimization within the saddle-point formulation. **Fig. 7(b)** shows the evolution of the identified flexoelectric coefficients, which converge to values very close to the ground truth (0.0848), namely $f_1 = 0.0849$ and $f_2 = 0.0872$. These results confirm the accuracy of the proposed approach in parameter identification. This performance demonstrates that incorporating the variational loss effectively overcomes the limitations of the deep energy method in inverse settings. A comparison with a baseline model trained without the variational loss, presented in **Supplementary Note 2**, shows that parameter convergence fails and the reconstructed fields become inaccurate, highlighting the necessity of this additional term for robust inverse identification. **Fig. 7(c)** further supports this by comparing the PINN-predicted displacement, electric potential, and strain fields with reference solutions obtained from mixed FEM. The high level of agreement across all quantities, along with minimal absolute errors, reaffirms the method's ability to simultaneously reconstruct field variables and estimate underlying flexoelectric parameters with high fidelity.

To further evaluate the proposed inverse framework, additional tests were conducted using different combinations of flexoelectric coefficients. Results for cases in which $f_1$ and $f_2$ are both negative or have opposite signs are presented in **Supplementary Note 3.**

## 5. Conclusion

This study proposed a deep energy-based PINN framework for solving both forward and inverse problems in linear flexoelectricity. By reformulating the governing equations as a saddle-point problem of the total potential energy, the approach facilitates a variational formulation that circumvents the direct computation of fourth-order derivatives, which often lead to poor convergence in conventional PINN. Finite element–based numerical integration is employed to evaluate the energy functional, and Dirichlet boundary conditions are enforced exactly via hard constraints using masking and auxiliary functions. The proposed method was validated against benchmark solutions obtained from a mixed finite element method for both direct and converse flexoelectric effects, with excellent agreement observed across displacement, electric potential, and strain fields. For inverse problems, a variational loss term based on the potential energy functional was introduced to enable the stable identification of unknown flexoelectric coefficients from partial electric potential data. The success of the framework in accurately reconstructing field responses and identifying material parameters demonstrates its robustness and flexibility for high-order coupled electromechanical systems. Although the current implementation relies on a structured mesh for numerical integration, the overall framework is readily extensible to more complex geometries, material systems, and experimental configurations, making it a promising tool for future physics-informed modeling in flexoelectricity and beyond.

**Declaration of Competing Interest**

The authors declare that they have no known competing financial interests or personal relationships that could have appeared to influence the work reported in this paper

**Data availability**

Data will be made available on request.

**Acknowledgements**

This work was supported by the National Research Foundation of Korea (NRF) grand funded by the Korea government (MSIT) (No. RS-2023-00222166 and No. RS-2023-00247245).

# Supplementary information

# Physics-informed neural network framework for solving forward and inverse flexoelectric problems

Hyeonbin Moon[1†], Donggeun Park[1†], Jinwook Yeo[1] and Seunghwa Ryu[1*]

[1]Department of Mechanical Engineering, Korea Advanced Institute of Science and Technology (KAIST), 291 Daehak-ro, Yuseong-gu, Daejeon 34141, Republic of Korea

[1†] Hyeonbin Moon and Donggeun Park contributed equally to this work.

[*]Corresponding author e-mail

ryush@kaist.ac.kr (Seunghwa Ryu)

**Supplementary Note 1. Masking and Auxiliary Functions for Enforcing Dirichlet Boundary Conditions**

The specific forms of the masking and auxiliary functions used in each case are summarized below:

(1) Forward problem: Direct flexoelectric

$$p_u(\boldsymbol{x}) = x, \qquad q_u(\boldsymbol{x}) = 0$$

$$p_\phi(\boldsymbol{x}) = y(1-y), \qquad q_\phi(\boldsymbol{x}) = 0.01y$$

(2) Forward problem: Converse flexoelectric

$$p_u(\boldsymbol{x}) = x, \qquad q_u(\boldsymbol{x}) = 0$$

$$p_\phi(\boldsymbol{x}) = y(1-y), \qquad q_\phi(\boldsymbol{x}) = 0$$

(3) Inverse problem

$$p_u(\boldsymbol{x}) = x(1-x)/10, \qquad q_u(\boldsymbol{x}) = 0$$

$$p_\phi(\boldsymbol{x}) = y(1-y)/10000, \qquad q_\phi(\boldsymbol{x}) = yNN_{m\ easure}(x)$$

In the inverse problem, since the electric potential data are not available in analytic form, a neural network surrogate $NN_{m\ easure}(x)$ is trained on the measured electric potential profile along the top edge of the beam. The accuracy of this surrogate model is validated in **Fig. S1**, confirming its high fidelity in representing the measured electric potential profile.

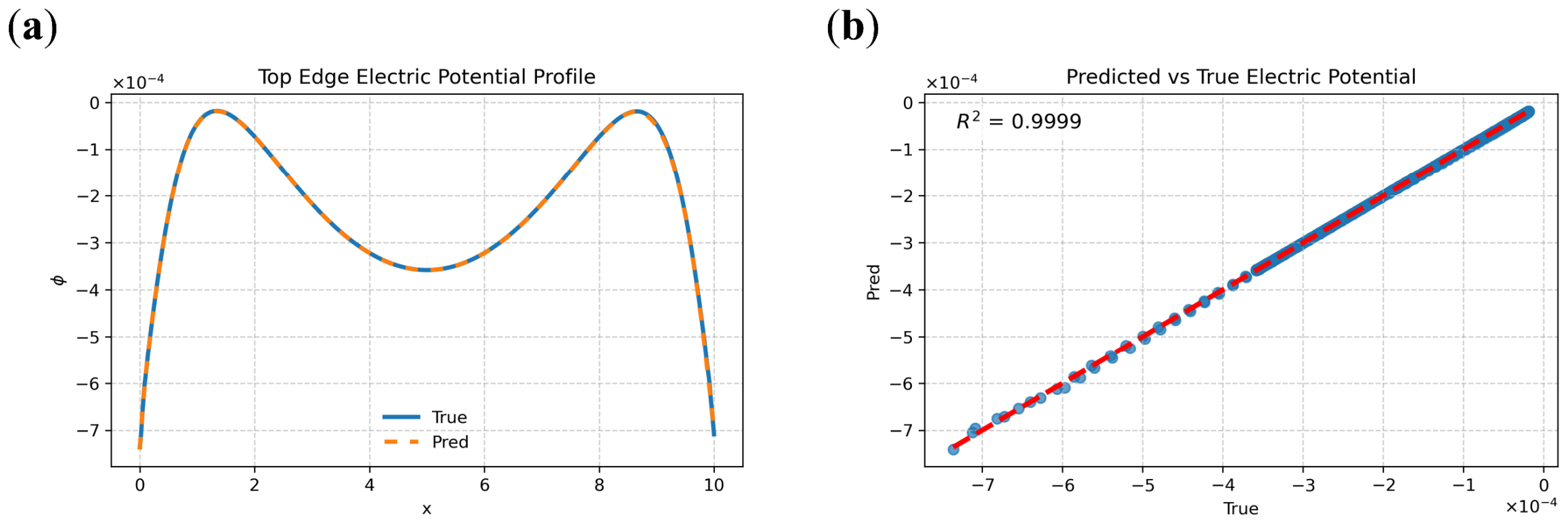

**Fig. S1. Accuracy of the neural network surrogate for the electric potential on top edge.** (a) Comparison of measured data (blue) and neural network prediction (orange). (b) $R^2$ plot between measured and predicted values.

**Supplementary Note 2. Inverse problem with and without Variational Loss**

Here, we investigate the effect of incorporating the variational loss term into the inverse formulation. While the main text demonstrates the effectiveness of the proposed PINN framework with this additional constraint, we further assess its contribution by comparing against baseline models trained without the variational term.

In the absence of a variational loss, the only available mechanism for identifying the flexoelectric coefficients $f_1$ and $f_2$ is through the extremization of the total potential energy. That is, the parameters are indirectly optimized by either minimizing or maximizing the total potential of the system. To systematically evaluate this limitation, we consider three distinct training scenarios:

- **Case 1**: Potential energy minimization

  The material parameters are optimized by minimizing the total potential energy.

- **Case 2**: Potential energy maximization

  The material parameters are optimized by maximizing the total potential energy.

- **Case 3**: Variational loss minimization

  The parameters are trained solely by minimizing a variational loss, without involving the total potential energy.

**Fig. S2** summarizes the inverse identification behavior under three training scenarios. In **Case 1**, all key metrics—including the identified flexoelectric coefficients, potential energies, and variational loss—diverge significantly during training, indicating unstable and non-convergent behavior. In **Case 2**, the parameters and potential energy appear to converge; however, the variational loss remains large, suggesting that the solution corresponds to a non-physical stationary point that does not satisfy the governing equations. Only **Case 3**, which incorporates the variational loss as part of the training objective, achieves consistent convergence across all metrics, confirming that the variational constraint is essential for obtaining physically meaningful solutions in inverse flexoelectric problems.

These results confirm that potential energy alone is insufficient for robust inverse identification, and that the variational loss plays a critical role in stabilizing training and guiding the optimization of flexoelectric coefficients.

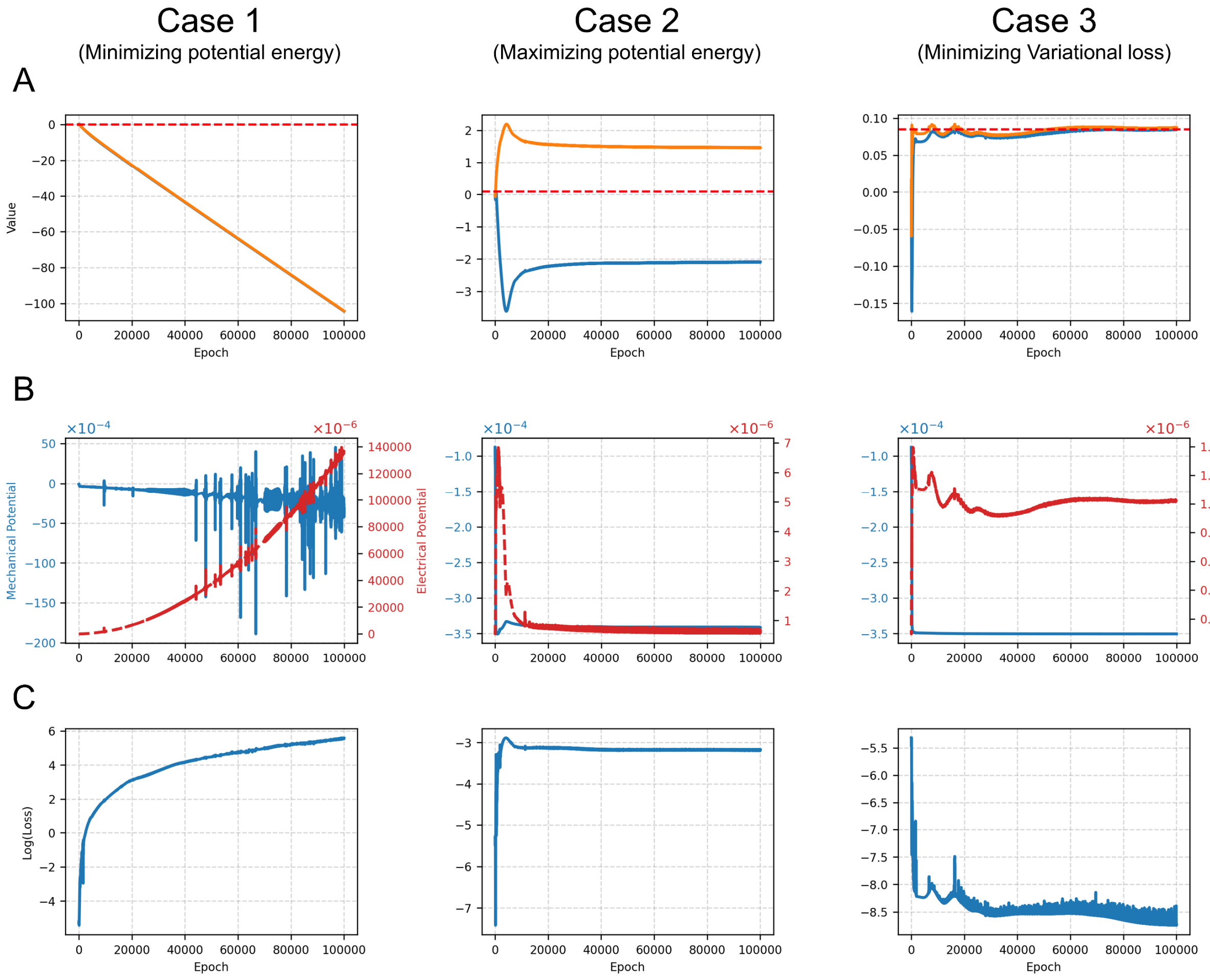

**Fig. S2. Training histories for inverse identification using three training schemes.** (a) Evolution of identified flexoelectric coefficients. (b) Mechanical and electrical potential energies. (c) Variational loss

**Supplementary Note 3. Inverse Identification of flexoelectric coefficients**

To further evaluate the robustness of the proposed inverse PINN framework, we conducted additional numerical experiments involving different combinations of flexoelectric coefficients. Specifically, two test cases were considered: $(i)$ $f_1 = 1 \times 10^{-6}, f_2 = -1 \times 10^{-6}$ $(i\ )$ $f_1 = -1 \times 10^{-6}, f_2 = -1 \times 10^{-6}$ .These values correspond to nondimensional coefficients of approximately $\mp 0.0848$ according to the reference scaling described in the manuscript. All results are presented in nondimensional form and summarized in **Fig. S3** and **Fig. S4**.

In case (i), the identified coefficients were $f_1 = 0.0850$ and $f_2 = 0.0850$; in case (ii), the estimates were $f_1 = -0.0832$ and $f_2 = -0.0833$, respectively Each figure includes: (a) the electric potential profile along the top edge used as input data; (b) the training history of the identified coefficients; and (c) the reconstructed displacement, electric potential, and strain fields in comparison with FEM reference solutions, along with absolute error maps. The accuracy of both parameter identification and field prediction across these varied scenarios highlights the reliability and adaptability of the proposed framework in inverse flexoelectric analysis.

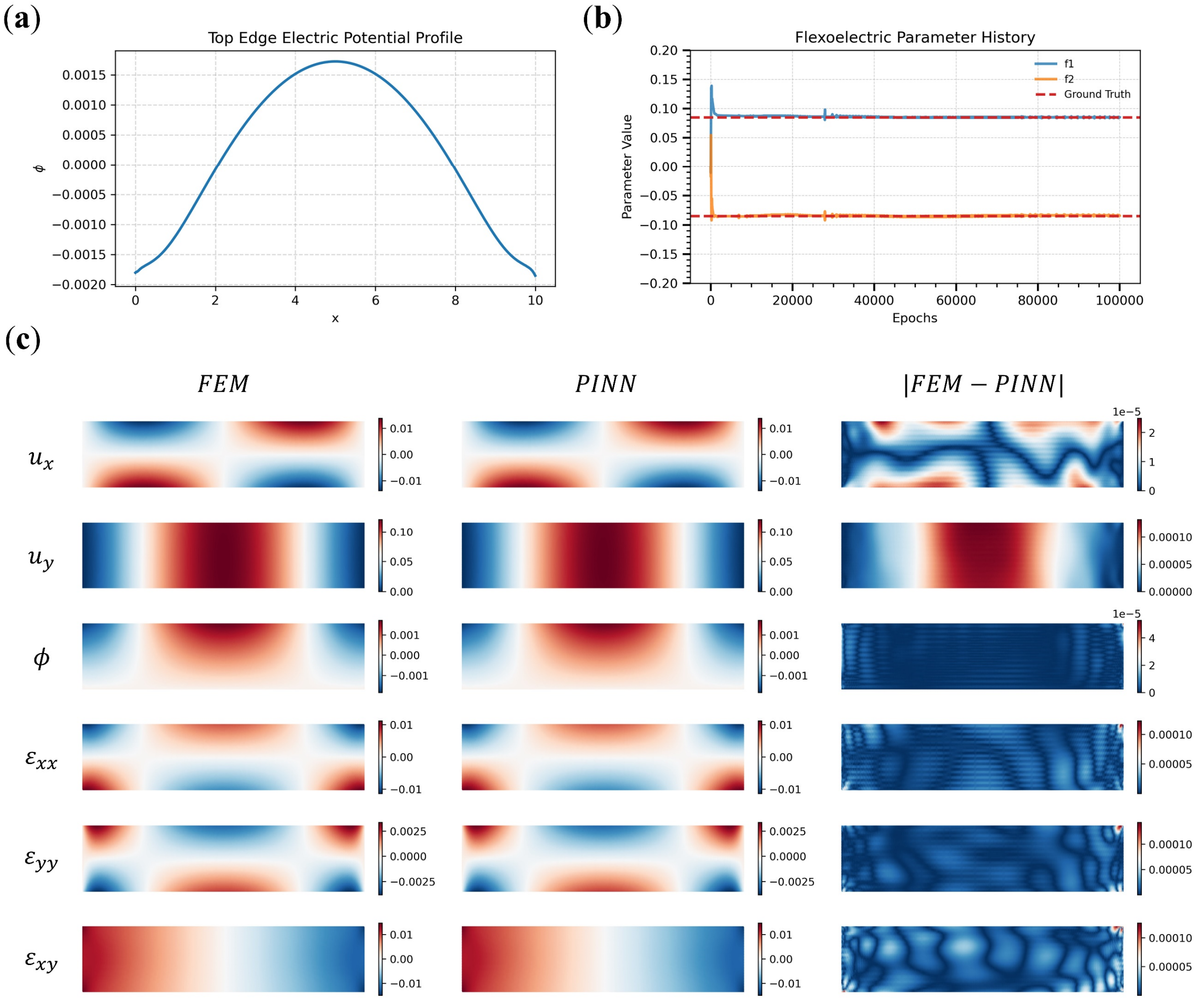

**Fig. S3. Inverse identification results for asymmetric flexoelectric coefficients $\boldsymbol{f_1 = 1 \times 10^{-6}, f_2 = -1 \times 10^{-6}}$.** (a) Measured electric potential along the top edge. (b) Training history of identified flexoelectric coefficients. (c) Comparison of PINN and FEM predictions: displacements, electric potential, strains, and absolute errors.

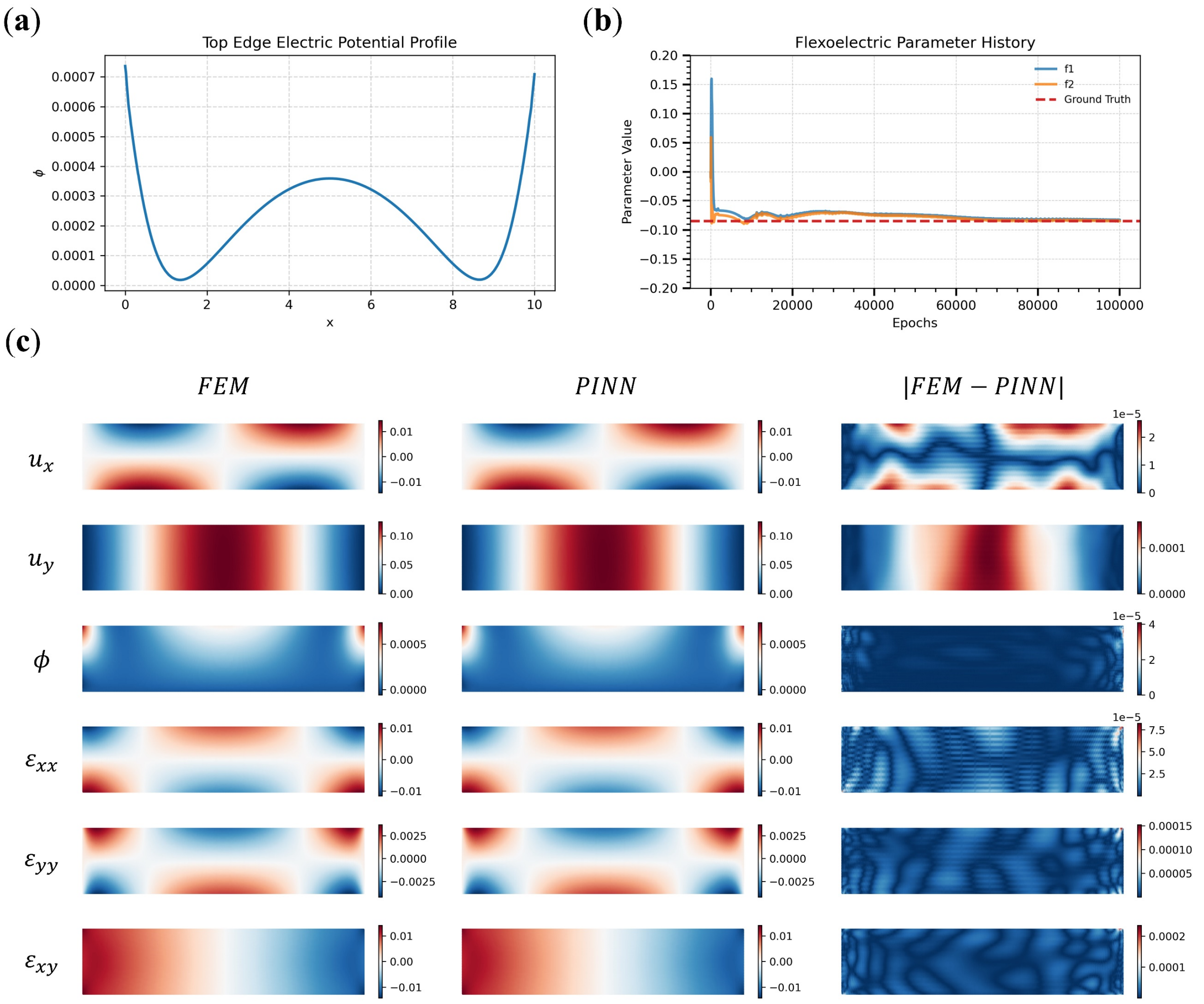

**Fig. S4. Inverse identification results for asymmetric flexoelectric coefficients $f_1 = -1 \times 10^{-6}, f_2 = -1 \times 10^{-6}$.** (a) Measured electric potential along the top edge. (b) Training history of identified flexoelectric coefficients. (c) Comparison of PINN and FEM predictions: displacements, electric potential, strains, and absolute errors.